\documentclass{ws-procs9x6-notrimmark}
\usepackage{times}
\usepackage{txfonts}
\usepackage{color}
\usepackage{subfigure}
\usepackage{ulem}
\newcommand{\ep}{\varepsilon}

\newcommand{\eqs}[1]{\begin{equation} \begin{split} #1\end{split} \end{equation} }

\newcommand{\ks}[1]{#1 \!\!\! \slash } 
\newcommand{\kks}[1]{#1 \!\! \slash } 

\newcommand{\ga}{\gamma^5}
\newcommand{\gmu}{\gamma^\mu}

\newcommand{\gmul}{\gamma_{\mu}}

\newcommand{\tra}{{\rm Tr}}

\newcommand{\etal}{{\it et al.}}

\newcommand{\I}{{\cal I}}
\newcommand{\Ip}{{\cal I}'}

\newcommand{\ce}[1]{Eq.~(\ref{#1})}

\newcommand{\cf}[1]{{Fig.~\ref{#1}}}

\newcommand{\nn}{\nonumber}

\newcommand{\beq}[1]{
\begin{equation}\label{#1}}
\newcommand{\eeq}{\end{equation}}
\newcommand{\bea}[1]{
\begin{eqnarray}\label{#1}}
\newcommand{\eea}{\end{eqnarray}}
\newcommand{\out}{\raise-3pt\hbox{\scriptsize    out}}

\begin{document}

\title{Single Transverse-Spin Asymmetry \\ in Hard-Exclusive Meson Electroproduction \\
in the Backward Region}

\author{\tiny  ~ \\ \small J.P. Lansberg$^1$\footnote{lansberg@cpht.polytechnique.fr}, B. Pire$^1$, L. Szymanowski$^2$}

\address{~ \\ $^1$Centre de Physique Th\'eorique, \'Ecole polytechnique, CNRS, 
F-91128, Palaiseau, France\\
$^2$Soltan Institute for Nuclear Studies, Warsaw, Poland
%\\
%$^*$E-mail: ab\_author@university.com\\
%www.university\_name.edu
}

\begin{abstract} \small
We discuss the relevance of studying  single transverse-spin asymmetry in 
hard-exclusive meson electroproduction in the backward region. Such an asymmetry could
help us discriminate between contributions from a soft baryon exchange in the $u$-channel 
and a hard parton-induced scattering.
\end{abstract}


\bodymatter

\section{Introduction}\label{aba:sec1}

In Ref.~\refcite{Pire:2005ax,Pire:2005mt,Lansberg:2007ec}, we introduced the framework to study
backward pion electroproduction, \eqs{\gamma^\star(q) N(p_1)  \to N'(p_2) \pi(p_\pi),}
on a proton (or neutron) target, in the Bjorken regime ($q^2$ large and $q^2/(2 p_1.q)$ fixed) 
in terms of a factorized amplitude (see \cf{fig:fact}) where a hard part is convoluted with  Transition Distribution Amplitudes (TDAs),  as well as the reaction,
\eqs{N(p_1) \bar N (p_2) \to \gamma^\star(q) \pi(p_\pi),}
in the  near forward region.\cite{Pire:2004ie,Lansberg:2007se,Lutz:2009ff}
 This extended the concept of 
Generalised Parton Distributions (GPDs). 
Such an extension of the GPD framework has already
been advocated in the pioneering work of Ref.~\refcite{Frankfurt:1999fp}. The TDAs involved in the description of Deeply-Virtual Compton Scattering (DVCS) 
in  the backward kinematics 
$ \gamma^\star (q) N(p_1)  \to N'(p_2) \gamma(p_\gamma)$
and the reaction
$N(p_1) \bar N (p_2) \to \gamma^\star(q) \gamma(p_\gamma)$ 
in the near forward region were given  in Ref.~\refcite{Lansberg:2006uh}.

Recently, a study of the TDAs in the meson-cloud 
model\cite{Pasquini:2006dv} has become available.\cite{Pasquini:2009ki}. Yet, more work is needed
before being able to proceed to {\it quantitative} comparisons between different
TDA models and between theory and experiments. For the time being, model
independent analyses -- looking for scaling or characteristic polarization effects -- sound more expedient. 
In this context, we would like to argue here that the study of target transverse-spin asymmetry 
(which we will denote SSA) could be used as a test of the 
dominance of a hard parton-induced scattering in the backward region at large $Q^2$ rather than that of a soft baryon exchange 
in the $u$-channel. Such a reaction would only generate phases through final state interactions, 
expected to decrease for $W^2 \gg (M+m_\pi)^2$ and large $Q^2$. On the contrary, 
 in reactions at the parton level, one expects an imaginary part to develop and to 
generate a SSA independently of whether $W^2$ and $Q^2$ are large or not.  This will be explained later on.

\begin{figure}[htb!]
\centering{
\includegraphics[height=6.5cm,clip=true]{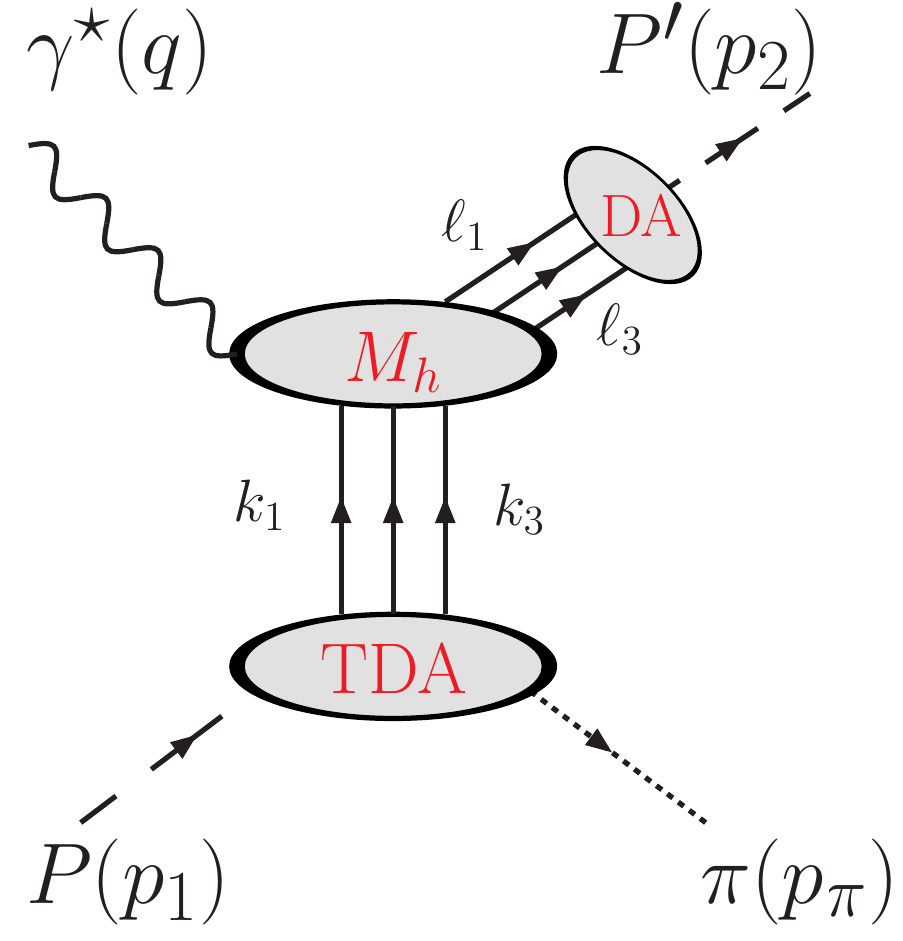}}
\caption{Illustration of the factorisation for backward electroproduction of a pion.}
\label{fig:fact}
\end{figure}

\begin{figure}[htb!]
\centering{
\includegraphics[width=\textwidth,clip=true]{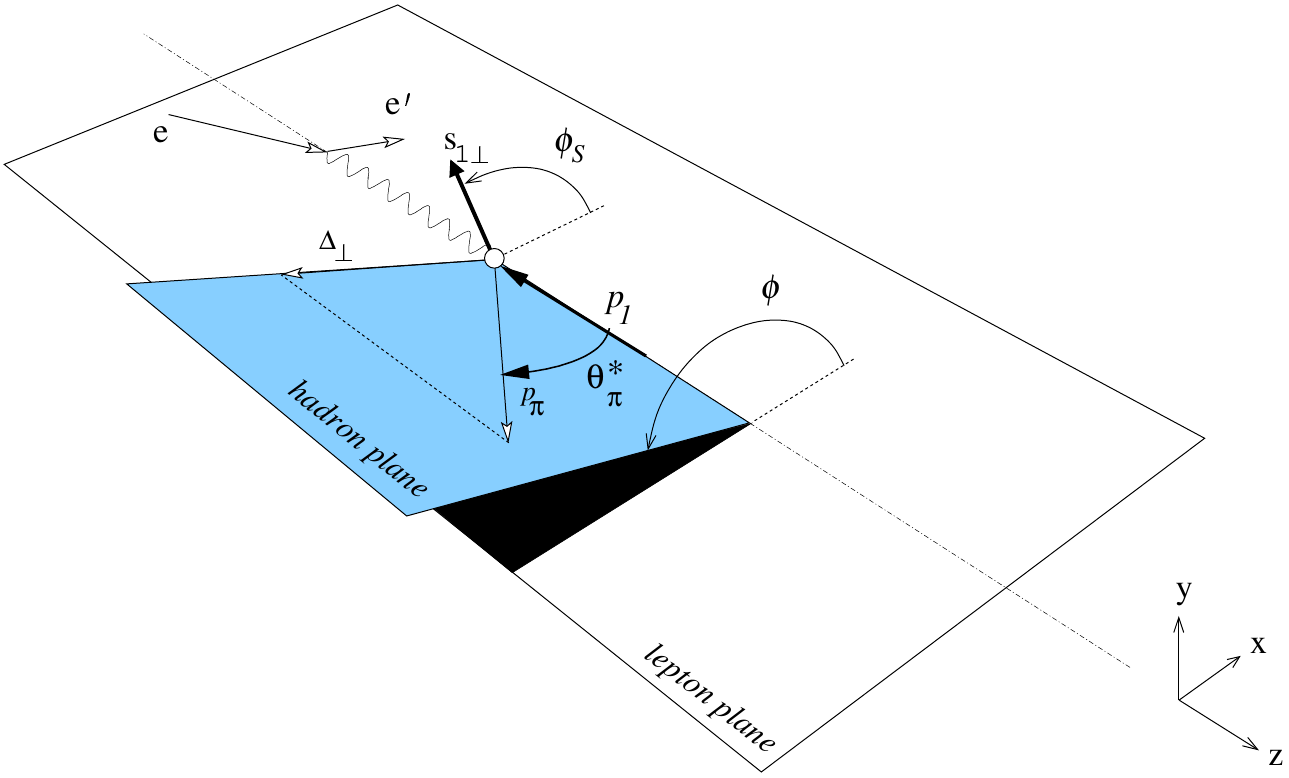}} 
\caption{ Kinematics of electroproduction of a pion and definition of the angles $\phi$ and $\phi_S$}
\label{fig:kin}
\end{figure}

\section{Some definitions}

The five-fold differential 
cross section for the process $e P \to e' P'\pi^0$ can be reduced
to a two-fold one -- expressible in the center-of-mass frame of the $P'\pi^0$ pair -- multiplied by a flux
factor $\Gamma$, $\frac{d^5\sigma}{dE_{e'}d^2\Omega_ed^2\Omega_{\pi}^\ast}=
\Gamma\ \frac{d^2\sigma}{d^2\Omega_{\pi}^\ast}$,
where $\Omega_e$ is the differential solid angle for the scattered electron
in the lab frame, and $\Omega_{\pi}^*$ is the differential solid angle for the
pion in the $P'\pi^0$ center-of-mass frame, such that $d\Omega_{\pi}^*=d\phi \, d \cos \theta^*_\pi$.
$\theta^*_\pi$ is defined as the polar angle between the virtual photon and the pion in the latter system (see \cf{fig:kin}).
$\phi$ is the azimuthal angle between the electron plane and the plane of the 
process $\gamma^\star P \to P' \pi^0$ (hadronic plane)
($\phi=0$ when the pion is emitted in the half plane containing the outgoing electron).

In general, we have contributions from different polarisations of the photon. For that
reason, we define four polarised cross sections, $d^2\sigma_{\mbox{\tiny T}}$,
$d^2\sigma_{\mbox{\tiny L}}$, $d^2\sigma_{\mbox{\tiny TL}}$ and
$d^2\sigma_{\mbox{\tiny TT}}$,\footnote{ For $L,x,y$, 
the linear polarisations of the virtual photon (for the definition of the $x$ \& $y$ axis, see \cf{fig:fact}), 
one defines\cite{Mulders:1990xw,Park:2007tn}
$d^2\sigma_{\mbox{\tiny L}}\propto {\cal M}^{{\tiny L}} ({\cal M}^{{\tiny L}})^\ast$, 
$d^2\sigma_{\mbox{\tiny T}}  \propto 1/2 [{\cal M}^{{\tiny x}} ({\cal M}^{{\tiny x}})^\ast
+{\cal M}^{{\tiny y}} ({\cal M}^{{\tiny y}})^\ast]$, $d^2\sigma_{\mbox{\tiny TL}}
\propto {\cal M}^{{\tiny x}} ({\cal M}^{{\tiny L}})^\ast
+{\cal M}^{{\tiny L}} ({\cal M}^{{\tiny x}})^\ast$ and $d^2\sigma_{\mbox{\tiny TT}}
\propto 1/2 [{\cal M}^{{\tiny x}} ({\cal M}^{{\tiny x}})^\ast
-{\cal M}^{{\tiny y}} ({\cal M}^{{\tiny y}})^\ast]$.
} which do not depend on $\phi$ but only
on $W$, $Q^2$ and $\theta^*_\pi$. The $\phi$ dependence is written as~\cite{Park:2007tn}
\eqs{
\frac{d^2\sigma}{d\Omega_{\pi}^*} =
\frac{d^2\sigma_{\mbox{\tiny T}}}{d\Omega_{\pi}^*}\ +
\ \epsilon\ \frac{d^2\sigma_{\mbox{\tiny L}}}{d\Omega_{\pi}^*}
+ \sqrt{2\epsilon(1+\epsilon)}\ \frac{d^2\sigma_{\mbox{\tiny TL}}}
{d\Omega_{\pi}^*}\cos{\phi} 
+\ \epsilon\ \frac{d^2\sigma_{\mbox{\tiny TT}}}{d\Omega_{\pi}^*}\cos{2\phi}.}
At the leading-twist accuracy, the QCD mechanism considered here contributes only 
to $\frac{d^2\sigma_{\mbox{\tiny T}}}{d\Omega_{\pi}^*}$. 

In the scaling regime, the amplitude for
$  \gamma^\star P(p_1)  \to P'(p_2) \pi(p_\pi) $
 in the backward kinematics -- namely small  $u=(p_\pi -p_1)^2=\Delta^2$ or $\cos \theta^*_\pi$ 
close to -1 -- involves the  TDAs $T(x_{i}, \xi, \Delta^2)$, 
where $x_i$ ($i=1,2,3$) denote the light-cone-momentum fractions carried by participant 
quarks and $\xi$ is the skewness parameter such that $2\xi =x_1+x_2+x_3$.

The amplitude is then a convolution of the proton DAs, a 
perturbatively-calculable-hard-scattering amplitude and the TDAs,
defined from the Fourier transform of a matrix element of a three-quark-light-cone operator between a 
proton and a meson state.
We have shown  that these TDAs obey QCD evolution equations,
which follow from the renormalisation-group equation of  the
three-quark operator. 
Their $Q^2$ dependence is thus completely  under control.

The momenta of the process $\gamma^\star P \to P' \pi $ are defined as in \cf{fig:fact}~\&~\ref{fig:kin}.
The $z$-axis is chosen along the initial nucleon and the virtual photon  momenta
and the $x-z$ plane is identified 
with the collision or hadronic plane (\cf{fig:kin}). Then, we define the 
light-cone vectors $p$ and $n$ ($p^2$=$n^2$=0) 
such that $2~p.n=1$, as well as $P=\frac{1}{2} (p_1+p_\pi)$, $\Delta=p_\pi -p_1$ and its 
transverse component $\Delta_T$ ($\Delta_T.\Delta_T=\Delta_T^2<0$). From those, we define  $\xi$ in an usual way as $\xi=-\frac{\Delta.n}{2P.n}$.

We  can then express the momenta of the particles through their  
Sudakov decomposition and, keeping the first-order corrections in the masses and $\Delta_T^2$, we have:
\eqs{\label{eq:decomp_momenta}
p_1&=\! (1+\xi) p + \frac{M^2}{1+\xi}n, q\simeq\!- 2 \xi \Big(1+ \frac{(\Delta_T^2-M^2)}{Q^2}\Big)  p + \frac{Q^2}{2\xi\Big(1+ \textstyle \frac{(\Delta_T^2-M^2)}{Q^2}\Big)} n,\\ \\
p_\pi&=\! (1-\xi) p +\frac{m_\pi^2-\Delta_T^2}{1-\xi}n+ \Delta_T,\Delta=\! - 2 \xi p +\Big[\frac{m_\pi^2-\Delta_T^2}{1-\xi}- \frac{M^2}{1+\xi}\Big]n
+ \Delta_T \nn}
\eqs{\label{eq:decomp_momentb}
p_2&\simeq- 2 \xi \frac{(\Delta_T^2-M^2)}{Q^2} p+\! \Big[\frac{Q^2}{2\xi\Big(1+ \textstyle \frac{(\Delta_T^2-M^2)}{Q^2}\Big)} -\frac{m_\pi^2-\Delta_T^2}{1-\xi}+ \frac{M^2}{1+\xi}\Big]n - \Delta_T,\\
}

For $\ep_x=(0,1,0,0)$ and $\ep_y=(0,0,1,0)$ with the axis definitions of \cf{fig:kin}, one may further 
specify that 
\eqs{\Delta_T=|\Delta_T| (\cos \phi \, \ep_x + \sin \phi \, \ep_y) \hbox{ and }
s_{T,1}=s_{1}= \cos \phi_S \, \ep_x + \sin \phi_S \, \ep_y,}
for the transverse spin of the target ($s_1.p_1=s_1.p=s_1.n=0$).

In the following, we use the same definition of the $p\to \pi^0$ TDAs as given by Eq. (15) of Ref. [\refcite{Lansberg:2007ec}].

 \section{Hard-amplitude calculation}

At leading order in $\alpha_s$, the amplitude ${\cal M}^\lambda_{s_1s_2}$  for 
$\gamma^\star(q,\lambda) P(p_1,s_1) \to P'(p_2,s_2) \pi^0(p_\pi)$ reads

\eqs{\label{eq:ampl-bEPM1}
{\cal M}^\lambda_{s_1s_2}=\underbrace{-i 
\frac{(4 \pi \alpha_s)^2 \sqrt{4 \pi \alpha_{em}} f_{N}^2}{ 54 f_{\pi}}}_{\cal C}&\frac{1}{Q^4}\Big[ 
\underbrace{ \bar u_2 \ks \ep(\lambda) \gamma^5 u_1}_{{\cal S}^\lambda_{s_1s_2}}
\underbrace{\int \!
\Bigg(2\sum\limits_{\alpha=1}^{7} T_{\alpha}+
\sum\limits_{\alpha=8}^{14} T_{\alpha}\Bigg)}_{{\cal I}}\\&-
\underbrace{ \bar u_2 \ks \ep(\lambda) \frac{\ks \Delta_{T}}{M} \gamma^5 u_1 }_{{\cal S'}^\lambda_{s_1s_2}}
\underbrace{\int \!
\Bigg(2\sum\limits_{\alpha=1}^{7} T'_{\alpha}+
\sum\limits_{\alpha=8}^{14} T'_{\alpha}\Bigg)}_{{\cal I'}}\Big]
,}
%\end{widetext}
where $\int \equiv \int\limits^{1+\xi}_{-1+\xi} \! \! \! dx_1dx_2dx_3\delta(2\xi -x_1-x_2-x_3)
 \int\limits_0^1 \! \! dy_1dy_2dy_3\delta(1-y_1-y_2-y_3)$, 
$u(p_1,s_1)\equiv~u_1$,  $\bar u(p_2,s_2)\equiv\bar u_2$  and the coefficients $T_{\alpha}$ and $T'_{\alpha}$$(\alpha=1,...,14)$ are functions of $x_{i}$,$y_{j}$,$\xi$ and $\Delta^2$ and  are given in Table 1 
of Ref. [\refcite{Lansberg:2007ec}]. 

The expression in \ce{eq:ampl-bEPM1} is to be compared with the leading-twist amplitude for the 
baryonic-form factor~\cite{CZ}
\eqs{\label{eq:ampl-FF} &{\cal M}^\lambda \propto  -i ( \bar u_2 \ks \ep(\lambda) u_1) \frac{\alpha_s^2 
f_{N}^2}{Q^4}  \int
\Bigg(2\sum\limits_{\alpha=1}^{7} T^p_{\alpha}(x_{i},y_{j},\xi,t)+
\sum\limits_{\alpha=8}^{14} T^p_{\alpha}(x_{i},y_{j},\xi,t)\Bigg)
.}

The factors $T^p_{\alpha}$ are very similar to the $T_{\alpha}$ obtained here. However, the integration domain is different. 
In the form factor case
\eqs{\int \hbox{stands for} \int \limits_{0}^{1} dx_1dx_2dx_3\delta(1 -\sum_i x_i) \int 
\limits_0^1  dy_1dy_2dy_3\delta(1-\sum_i y_i).}
Consequently, the integration of denominators in $T^p_\alpha$ such as $1/(x_i+i \varepsilon)$ do not generate any
imaginary parts. On the contrary, the integrations of similar denominators in $T_\alpha$ and $T'_\alpha$ over the
 TDA integration domain will
generate an imaginary part when passing from the ERBL region (all $x_i>0$) to one of the DGLAP regions (one $x_i<0$).
This will be the source of the SSA as we will show later on. 

\section{Single Transverse Spin Asymmetry}

In order to study a possible SSA we shall study the quantity $\sigma^{s_1}-\sigma^{-s_1}$  with
the definition
\eqs{\sigma^{s_1}=\sum_\lambda\sum_{s_2} ({\cal M}^\lambda_{s_1s_2})
({\cal M}^\lambda_{s_1s_2})^\ast.}

Since we are interested in the leading twist contribution of this asymmetry, we can sum only over the transverse
polarisation of the virtual photon using 
$\underset{\lambda=x,y}{\sum} \ep(\lambda)^\mu (\ep(\lambda)^\nu)^\ast=-g^{\mu\nu}+(p^\mu n^\nu+p^\nu n^\mu)/(p.n)$.
 The sum on the final-proton spin $s_2$ is done using 
$ \underset{s_2}{\sum}u_\alpha(p_2,s_2) \bar u_\beta(p_2,s_2)=(\ks p_2+M)_{\alpha \beta}$. As regards the 
initial-proton spinor, one uses the following relation involving its  transverse spin $s_1$, 
$u_\alpha(p_1,s_1) \bar u_\beta(p_1,s_1)=1/2 (1+\ga \kks s_1)(\ks p_1+M)_{\alpha \beta}$.

Hence, one has 
\eqs{
\sum_{\lambda=x,y}\sum_{s_2} 
({\cal M}^\lambda_{s_1s_2})&
({\cal M}^\lambda_{s_1s_2})^\ast=
 \frac{|{\cal C}|^2}{Q^8}\Big(-g^{\mu\nu}+\frac{p^\mu n^\nu+p^\nu n^\mu}{p.n}\Big)\times\\
 \Big[
\bar u_2 \gmu \ga u_1 \I 
&-\bar u_2 \gmu \frac{\ks \Delta_T}{M} \ga u_1 \Ip 
\Big]
\times \Big[
-\bar u_1  \ga \gmul u_2 \I^\ast 
+\bar u_1 \ga \frac{\ks \Delta_T}{M} \gmul u_1 \Ip^\ast 
\Big]\\&~~~~~~~~~~~
=
 \frac{|{\cal C}|^2}{Q^8}\Big(-g^{\mu\nu}+\frac{p^\mu n^\nu+p^\nu n^\mu}{p.n}\Big)\times \\\Big[
-&\tra ((\ks p_2 +M) \gmu \ga (\ks p_1 +M) \frac{(1+\ga \kks s_1)}{2} \ga \gmul) \I \I^\ast \\+&
\tra ((\ks p_2 +M) \gmu \frac{\ks \Delta_T}{M} \ga (\ks p_1 +M) \frac{(1+\ga \kks s_1)}{2} \ga \gmul) \I' \I^\ast \\+&
\tra ((\ks p_2 +M) \gmu \ga (\ks p_1 +M) \frac{(1+\ga \kks s_1)}{2} \ga \frac{\ks \Delta_T}{M} \gmul) \I \I'^\ast \\-&
\tra ((\ks p_2 +M) \gmu \frac{\ks \Delta_T}{M} \ga (\ks p_1 +M) \frac{(1+\ga \kks s_1)}{2} \ga \frac{\ks \Delta_T}{M} \gmul) \I' \I'^\ast 
\Big]
}

Dropping the contributions proportional to the proton mass, the spin asymmetry reads ($\epsilon_{0123}=-\epsilon^{0123}=+1$)
\eqs{\sigma^{s_1}-\sigma^{-s_1}=& 8  \frac{|{\cal C}|^2}{Q^6} \frac{1+\xi}{\xi} \frac{\epsilon^{n p s_1 \Delta_T}}{M}  \Im m(\Ip \I^\ast)\\
=&  - 4 \frac{|{\cal C}|^2}{Q^6} \frac{|\Delta_T|}{M} \frac{1+\xi}{\xi}  \sin(\phi-\phi_S) \Im m(\Ip \I^\ast).
}
Comparing with the expressions for the unpolarised cross section obtained in Ref.~\refcite{Lansberg:2007ec}, one concludes that
the asymmetry for the hard-parton induced contribution is leading-twist as soon as $\Delta_T\neq 0$ and $\I$ or $\Ip$ are no longer pure real or pure 
imaginary numbers. This is precisely what one expects when DGLAP contributions are taken into account\cite{inprogress}.

\section{Discussion and conclusion}

Although the knowledge of {\it baryon to meson} TDAs has recently improved significantly thanks to 
a first study in the meson cloud model~\cite{Pasquini:2009ki} and another one focused on their spectral 
representations\cite{Pire:2010if}, model-independent observables aimed at studying the backward regime of
meson electroproduction will still be the bread-and-butter of this field for the months to come. 

In this context, we find it particularly relevant to emphasize that the study of the asymmetry of the target transverse  
spin would reveal unique information on the nature of the particles exchanged in the $u$ channel, be it 
a ``mere'' baryon slightly off-shell, or three perturbative quarks. For non-vanishing transverse momenta ($\Delta_T$),
 one expects in the latter case an asymmetry of the same order as the unpolarised cross section, while, in the former
case, they would be most likely decreasing for increasing $W^2$ and $Q^2$.

\section*{Acknowledgments}

We thank S.J. Brodsky, 
G. Huber, V. Kubarovsky, K.J. Park, B.~Pasquini, K. Semenov, P. Stoler for useful and motivating discussions as well as
the organizers of the workshop ``Exclusive 2010'' for their kind invitation to present our work.
This work is partly supported by the ANR contract BLAN07-2-191986. L.Sz. acknowledges the support 
by the Polish Grant N202 249235.

%%%%%%%%%%%%%%%%%%%%%%%%%%%%%%%%%%%%%%%%%%%%%%%%%%%%%%%%%%%%%%%%%%%%%%%%%%%


\begin{thebibliography}{99}

%\bibitem{TDApiproton}
%\cite{Pire:2005ax}
\bibitem{Pire:2005ax}
  B.~Pire and L.~Szymanowski,
  %``QCD analysis of anti-p N $\to$ gamma* pi in the scaling limit,''
  Phys.\ Lett.\ B {\bf 622} (2005) 83
  [arXiv:hep-ph/0504255]. 
  %%CITATION = HEP-PH 0504255;%%



%\cite{Pire:2005mt}
\bibitem{Pire:2005mt}
  B.~Pire and L.~Szymanowski,
  %``A QCD analysis of anti-p N --> gamma* pi and anti-p N --> gamma* gamma.
  %Where is the pion in the proton?,''
  PoS {\bf HEP2005} (2006) 103
  [arXiv:hep-ph/0509368].
  %%CITATION = HEP-PH 0509368;%%

%\cite{Lansberg:2007ec}
\bibitem{Lansberg:2007ec}
  J.~P.~Lansberg, B.~Pire and L.~Szymanowski,
  %``Hard exclusive electroproduction of a pion in the backward region,''
  Phys.\ Rev.\  D {\bf 75} (2007) 074004
  [Erratum-ibid.\  D {\bf 77} (2008) 019902]
  [arXiv:hep-ph/0701125]. Note that the differential cross-section plots are in nb/sr, not pb/sr as indicated.
  %%CITATION = PHRVA,D75,074004;%%


%\cite{Pire:2004ie}
\bibitem{Pire:2004ie}
%\bibitem{TDApigamma}
  B.~Pire and L.~Szymanowski,
%   ``Hadron annihilation into two photons and backward VCS in the scaling
  %regime of QCD,''
  Phys.\ Rev.\ D {\bf 71} (2005) 111501
  [arXiv:hep-ph/0411387].
  %%CITATION = HEP-PH 0411387;%%



%\cite{Lansberg:2007se}
\bibitem{Lansberg:2007se}
  J.~P.~Lansberg, B.~Pire and L.~Szymanowski,
  %``Production of a pion in association with a high-Q2 dilepton pair in
  %antiproton-proton annihilation at GSI-FAIR,''
  Phys.\ Rev.\  D {\bf 76} (2007) 111502(R)
  [arXiv:0710.1267 [hep-ph]].
  %%CITATION = PHRVA,D76,111502;%%

%\cite{Lutz:2009ff}
\bibitem{Lutz:2009ff}
  M.~F.~Lutz, \etal~ [The PANDA
                  Collaboration],
  %``Physics Performance Report for PANDA: Strong Interaction Studies with
  %Antiprotons,''
  arXiv:0903.3905 [hep-ex].
  %%CITATION = ARXIV:0903.3905;%%





%\cite{Frankfurt:1999fp}
\bibitem{Frankfurt:1999fp}
  L.~L.~Frankfurt, P.~V.~Pobylitsa, M.~V.~Polyakov and M.~Strikman,
  %``Hard exclusive pseudoscalar meson electroproduction and spin structure  of
  %a nucleon,''
  Phys.\ Rev.\ D {\bf 60} (1999) 014010
  [arXiv:hep-ph/9901429];
  %%CITATION = HEP-PH 9901429;%%
%\cite{Frankfurt:2002kz}
%\bibitem{Frankfurt:2002kz}
  L.~Frankfurt, M.~V.~Polyakov, M.~Strikman, D.~Zhalov and M.~Zhalov,
  %``Novel hard semiexclusive processes and color singlet clusters in hadrons,''
 in {\it  Newport News 2002, Exclusive Processes at High Momentum Transfer} (edited by A. Radyushkin, P. Stoler; Singapore, World Scientific, 2002, pp 361-368),
 arXiv:hep-ph/0211263.
  %%CITATION = HEP-PH/0211263;%%



%\cite{Lansberg:2006uh}
\bibitem{Lansberg:2006uh}
  J.~P.~Lansberg, B.~Pire and L.~Szymanowski,
  %``Backward DVCS and proton to photon transition distribution amplitudes,''
  Nucl.\ Phys.\ A {\bf 782} (2007) 16
  [arXiv:hep-ph/0607130].
  %%CITATION = HEP-PH 0607130;%%


%\cite{Pasquini:2006dv}
\bibitem{Pasquini:2006dv}
  B.~Pasquini and S.~Boffi,
  %``Virtual meson cloud of the nucleon and generalized parton distributions,''
  Phys.\ Rev.\ D {\bf 73} (2006) 094001
  [arXiv:hep-ph/0601177].
  %%CITATION = HEP-PH 0601177;%%

%\cite{Pasquini:2009ki}
\bibitem{Pasquini:2009ki}
  B.~Pasquini, M.~Pincetti and S.~Boffi,
  %``Parton content of the nucleon from distribution amplitudes and transition
  %distribution amplitudes,''
  Phys.\ Rev.\  D {\bf 80} (2009) 014017
  [arXiv:0905.4018 [hep-ph]].
  %%CITATION = PHRVA,D80,014017;%%


%\cite{Pobylitsa:2001cz}
\bibitem{Pobylitsa:2001cz}
  P.~V.~Pobylitsa, M.~V.~Polyakov and M.~Strikman,
  %``Soft pion theorems for hard near threshold pion production,''
  Phys.\ Rev.\ Lett.\  {\bf 87} (2001) 022001
  [arXiv:hep-ph/0101279].
  %%CITATION = HEP-PH 0101279;%%



%\cite{Mulders:1990xw}
\bibitem{Mulders:1990xw}
  P.~J.~Mulders,
  %``MODIFICATIONS OF NUCLEONS IN NUCLEI AND OTHER CONSEQUENCES OF THE QUARK
  %SUBSTRUCTURE,''
  Phys.\ Rept.\  {\bf 185} (1990) 83.
  %%CITATION = PRPLC,185,83;%%

%\cite{Park:2007tn}
\bibitem{Park:2007tn}
  K.~Park {\it et al.}  [CLAS Collaboration],
  %``Cross sections and beam asymmetries for $\vev{e}p \to en\pi^+$ in the
  %nucleon resonance region for $1.7 \le Q^2 \le 4.5 (GeV)^2$,''
  Phys.\ Rev.\  C {\bf 77} (2008) 015208
  [arXiv:0709.1946 [nucl-ex]].
  %%CITATION = PHRVA,C77,015208;%%



\bibitem{CZ}
V.~L.~Chernyak and A.~R.~Zhitnitsky,
%``Asymptotic Behavior Of Exclusive Processes In QCD,''
Phys.\ Rept.\  {\bf 112}, 173 (1984);
%%CITATION = PRPLC,112,173;%%


%\cite{Pire:2010if}
\bibitem{Pire:2010if}
  B.~Pire, K.~Semenov-Tian-Shansky and L.~Szymanowski,
  %``A spectral representation for baryon to meson and baryon to photon
  %transition distribution amplitudes,''
  arXiv:1008.0721 [hep-ph].
  %%CITATION = ARXIV:1008.0721;%%


\bibitem{inprogress}
J.P. Lansberg, \etal, work in progress.
%


\end{thebibliography}
\end{document}